\documentclass[twocolumn]{emulateapj}

\bibpunct{(}{)}{;}{a}{}{,}

\newcommand{\msun}{\mbox{$M_\odot$}}
\newcommand{\rsun}{\mbox{$R_\odot$}}
\def\be{\begin{eqnarray}}
\def\ee{\end{eqnarray}}
\def\lsim{\mathrel{\rlap{\lower3pt\hbox{\hskip1pt$\sim$}}
     \raise1pt\hbox{$<$}}} 
\def\gsim{\mathrel{\rlap{\lower3pt\hbox{\hskip1pt$\sim$}}
     \raise1pt\hbox{$>$}}} 

\shorttitle{Hypercritical Accretion in M33 X-7}
\shortauthors{Moreno M\'endez et al.}

\begin{document}

\title{The Case for Hypercritical Accretion in M33 X$-$7}


\author{Enrique Moreno M\'endez, Gerald E. Brown}
\affil{Department of Physics and Astronomy,
               State University of New York, Stony Brook, NY 11794, USA.}

\email{EMM: moreno@grad.physics.sunysb.edu}

\email{GEB: gbrown@insti.physics.sunysb.edu}

\author{Chang-Hwan Lee}
\affil{Department of Physics, Pusan National University,
              Busan 609-735, Korea.}
\email{CHL: clee@pusan.ac.kr}

\and

\author{Il H. Park}
\affil{Research Center of MEMS Space Telescope, Dept. of Physics, Ewha Womans University, Seoul 120-750, Korea}
\email{IHP: ipark@ewha.ac.kr}


\begin{abstract}
The spin parameter of the black hole in M33 X$-$7 has recently been measured to be $a_\star=0.77\pm0.05$ \citep{Liu08}.  It has been proposed that the spin of the $15.65\msun$ black hole is natal.  We show that this is not a viable evolutionary path given the observed binary orbital period of 3.45 days since the explosion that would produce a black hole with the cited spin parameter and orbital period would disrupt the binary.  Furthermore, we show that the system has to be evolved through the hypercritical mass transfer of $\sim5\msun$ from the secondary star to the black hole.
\end{abstract}


\section{Introduction}\label{intro}

The black-hole spin in M33 X$-$7 has been very accurately measured to be
\[
a_\star=0.77\pm0.05
\]
in dimensionless spin parameter \citep{Liu08}.  The authors of this paper show that {\it ``In order to achieve a spin of $a_\star=0.77$ via disk accretion , an initially non-spinning  black hole must accrete $4.9\msun$ from its donor in becoming the $M_{BH}=15.65\msun$ that we observe today.  However to transfer this much mass even in the case of Eddington limited accretion ($\dot{M}\sim4\times10^{-8}\msun/$yr) requires $\sim120$ million years whereas the age of the system is only $2-3$ million years.  Thus the spin of M33 X$-$7 must be natal, which is the same conclusion that has been reached for two other stellar black holes"} (\citet{Sha06} and \citet{McC06}). However, \citet{Liu08} noted that their spin derivation is model-dependent and subject to possible systematic errors.

\citet{Lee02} predicted the spin parameters of Nova Sco (X-ray Nova Scorpii 1994) and Il Lupi (4U 1543$-$47) to be $\sim0.8$, with small effects after they were born in the explosion from mass accretion; i.e., predicted them as natal.  However \citet{BLMM07} showed that the rotational energy in such binaries scaled inversely with the donor mass at the time of common-envelope evolution preceding the explosion in which the black hole was born.  The donor masses of Nova Sco and Il Lupi are $\sim2\msun$ whereas that of M33 X$-$7 was $\sim80\msun$, so \citet{BLMM07} suggested that the $3.45$ day period of M33 X$-$7 resulted from a dark explosion; the high spin parameter would have resulted from mass accretion.  This mass accretion would have had to take place at hypercritical rate as we discuss in this letter.

We briefly comment on the hypercritical accretion for $\dot{M}/\dot{M}_{Edd}\sim10^3$ or greater \citep{Bro94}.  The scenario begins with Bondi accretion through the sonic point, which is often greatly larger than accretion at the Eddington limit.  Because it had been worked out for a value of $0.31\times10^4\dot{M}_{Edd}$ by \citet{Bro94} and because this value is in the middle of those we shall use in stellar evolution, we use this value, although it could be much greater.  The \citet{Bro94} work had been carried out earlier in all detail by \citet{Hou91} and checked by \citet{Che95}.
We note that there is still considerable controversy in the astrophysical community about whether hypercritical accretion can take place or not.
However, the general point we address is that if $\dot{M}$ exceeds $\dot{M}_{Edd}$, then some of the accretion energy must be removed by means other than photons.  In the case of hypercritical accretion, this excess energy can be carried off by neutrino pairs \citep{Bro94}.
In the case of a neutron star, neutrino losses allow the matter flow to join smoothly onto the neutron star surface.  In the case of a black hole, the neutrino losses let the matter flow smoothly over the event horizon and disappear into the black hole.

In the work of \citet{Pod03}, we note two possible stages where hypercritical ($\dot M/\dot M_{Edd} \gsim 10^3$) or supercritical accretion may take place.  \citet{Pod03} evolve a binary with $M_{BH}=12\msun$, $m_{secondary}=25\msun$ and orbital period of 6.8 days.
\begin{itemize}
\item[a)] Hypercritical accretion stage: They show that after the formation of the black hole, (first scenario) the secondary star will overflow its Roche Lobe and will transfer mass at a rate which peaks at $3\times10^{-3}\msun$yr$^{-1}$. The binary detaches after about $10^4$ years, once the secondary has been reduced to the mass of the black hole (or earlier if the stellar wind is strong enough).
\item[b)] Supercritical accretion stage: Mass transfer continues at a lower rate of $\sim3\times10^{-6}\msun$yr$^{-1}$ for up to another few $10^{6}$ years through a directed wind.
\end{itemize}
Nevertheless \citet{Pod03} restrict accretion into the black hole to the Eddington limit.  With this same assumption they were able to get GRS 1915$+$105 up to a spin parameter $a_\star=0.9$.  However, \citet{McC06} measured its spin parameter to be $a_\star\sim0.98-0.99$.
We were able to get the spin parameter up to the measured $a_\star>0.98$ with hypercritical accretion \citep{BLMM07} (see the discussion on P.355. of \citet{Bet03}).

If we suppress the assumption that the rate of accretion is limited to the Eddington limit we observe that the binary Cyg X$-$1 in \citet{Pod03} would be able to transfer up to $\sim30\msun$ during the first thermal timescale (assuming there is that much mass in the system), and another few solar masses afterwards, during the period where the black hole and the secondary star are detached, before the secondary fills again its Roche lobe during the red giant stage.

Cyg X$-$1, V4641 Sgr and GRS 1915$+$105 are similar in that the donors in all cases were more massive than the black hole at the time the black hole was formed.  The hypercritical accretion for GRS 1915$+$105 is necessary to bring $a_\star$ up to $a_\star>0.98$ \citep{BLMM07}.  For the purposes of discussing Cyg X$-$1 including the hypercritical nature of the accretion the detailed evolution of \citet{Pod03} is useful.  Given hypercritical accretion, M33 X$-$7 can be straightforwardly discussed in a similar way as we show in the next sections.

In Sec.~\ref{M33X7}, we discuss what would be the consequences if the current spin of M33 X-7 were natal. We discuss a few problems in this scenario. In Sec.~\ref{Evol} we discuss the case for hypercritical accretion in M33 X-7 as an alternative way of making high spin of black hole in M33 X-7. We summarize our conclusion in Sec.~\ref{Conc}.

\section{Consequences in M33 X$-$7}\label{M33X7}

In this section, we ask for M33 X$-$7 what the consequences would be were the currently observed spin of the black hole all natal.

Most important for the binary evolution is that the helium-star (progenitor of the black hole) is spun up by the secondary star so that these ``helium-stars will be fully synchronized with their orbital motion throughout their core-helium burning"; i.e., there is tidal locking of the helium-star with the secondary star \citep{Heu07}.
Hence, the spin of helium-star and the orbital motion of binary being locked together, and the angular momentum of He-star is transferred to that of black hole as the helium star falls into the latter \citep{Lee02}).
In that case, with the currently measured spin parameter $a_\star=0.77$,  the preexplosion orbital period of M33 X-7 would be essentially the same as for Nova Sco, which  \citet{Lee02} predicted to be 0.4 days with spin parameter $\sim0.75$ (see Figure 12 of \citet{Lee02}). This prediction was confirmed by
\citet{Sha06} with the measurement of $a_\star=(0.65-0.75)$ for Nova Sco.

Here we summarize a few problems with this scenario.
\begin{itemize}
\item[a)]
This tidal locking leads to a strange and complex situation for M33 X-7. Because of the short $0.4$ day period, the helium star squashes inside of the $\sim70\msun$ star, with orbital separation $a\sim 10\rsun$ (McClintock, private communication).  The tidal locking should be more effective in this case because these stars are much more massive and close together than those considered by \citet{Heu07}.

\item[b)]
With high spin angular momentum of helium-star (black hole progenitor),
the black hole is born in the Blaauw-Boersma explosion,\footnote{For a detailed description of Blaauw-Boersma Kick, and a source of the relevant formula~\ref{P1-P2} see the appendix in \citet{Bro01}} in which black hole binary can have system velocity due to the sudden mass loss during the explosion. With given preexplosion orbital period $\sim 0.4$ day and the present one of 3.45 days, M33 X-7 should have lost more than half of the system mass and couldn't survive the explosion if there were no hypercritical accretion as we discuss below.

\end{itemize}

In the case of Nova Sco, the explosion involved a mass loss of several
solar masses \citep{Nel99}.
The heliocentric radial system velocity of Nova Sco is $-150\pm19$km s$^{-1}$.  After correction for peculiar motion of the sum and differential Galactic rotation, the magnitude of the velocity stands out as being higher than any other dynamically identified Galactic black hole candidate \citep{Bra95}.
Given the donor mass of $\sim2\msun$ and the black hole mass of $(5.1-5.7)\msun$,
it lost nearly half of its system mass in the explosion \citep{Nel99}.
The reason why Nova Sco is the most energetic explosion among the soft X-ray transient sources is that the explosion energy has to be big enough to expel nearly half of its system mass. We believe that this energy was provided by the black hole spin.
The present remaining rotational energy is $430$ bethes (1 bethe $=10^{51}$ ergs).
\citet{Lee02} found that in Nova Sco most of the rotational energy is natal.

Given the same spin parameter in the natal spin of M33 X$-$7, it would have $\sim 3$ times more rotational energy than Nova Sco, because of the $\sim 3$ times more massive black hole, about half of $\msun c^2$!
In between the explosion and the present time, no forces act on the binary assuming the (negligible) sub-Eddington rate of accretion.  In other words, the explosion must convert the originally $0.4$ day period into the present one of $3.45$ days.  We take the black hole mass after the explosion to be the present one, since accretion at the Eddington limit changes its mass negligibly in 2-3 million years.

In the Blaauw-Boersma explosion, assuming rapid circularization,
\be
P_2=\left(1+\frac{\Delta M}{M_{BH}+m}\right)^2 P_1
\label{P1-P2}
\ee
where the black-hole mass $M_{BH}=M_{He}-\Delta M$ with the mass loss $\Delta M$ during the explosion, $m$ is the mass of the secondary star at the time of the explosion, $P_1$ the pre-explosion period, and $P_2$ is the post-explosion period.  It is well known that once the mass loss is half of the system mass ($\Delta M_{\rm breakup} = M_{BH}+m$) the binary becomes unstable and breaks up.  This happens at
\be
\left(\frac{P_{breakup}}{P_1}\right)=\left(1+\frac{\Delta M_{\rm breakup}}{M_{BH}+m}\right)^2=4.
\ee
With  $P_1 \sim 0.4$ days for $a_\star=0.77$,
the breakup period is 1.6 days, less than half the present $3.45$ days observed in M33 X$-$7; i.e., the binary would break up during the explosion.  Thus, there must be less mass loss in the explosion and mass must be transferred from the secondary star to the black hole following the explosion, as discussed by \citet{Bet03}, in order to achieve the observed spin parameter.

From the above consideration, we believe that the present value of spin parameter $a_\star=0.77$ for M33 X-7 cannot be the natal one.

Nova Sco is completely different in the respect that the companion is $\sim 2\msun$, much lighter than that in M33 X-7, and the most of the black hole spin energy is natal.
In Nova Sco, with the same natal period of $P_1\simeq0.4$ days, the breakup period should be the same $P_{breakup}=1.6$ days. The $\sim2\msun$ secondary star in Nova Sco is some billions of years old, so that even with accretion limited by Eddington, it could have transferred appreciable mass to the black hole.  \citet{Lee02} found this to be $0.41\msun$ which, if conservative, would have increased the period of Nova Sco following the explosion to be $1.5$ days.  It is the proximity of this period to the $1.6$ days (breakup period) which makes the system velocity of Nova Sco to be higher than any other Galactic black hole candidate.

\section{Evolution of M33 X$-$7}\label{Evol}

In the previous section, we have discussed that the present value for the spin parameter cannot be the natal one. In addition, from Figure 12 of \citet{Lee02}, we see that the 3.45 day period corresponds to a low natal spin parameter $a_\star\sim 0.12$ which is much lower than the observed one $a_\star=0.77$. So we believe that the spin up of the black hole has to be caused by the accretion from the companion.
Knowing that the present day orbital period and spin parameter of the black hole in M33 X-7, we can estimate from Figure 6 of \citet{Bro00} that about $(40-50)\%$ of the mass of the black hole had to be accreted after its formation.

Now we can obtain the orbital period before the accretion in M33 X$-$7 assuming conservative mass transfer,
\be
P_3&=&\left(\frac{M_{BH,4}\times m_4}{M_{BH,3}\times m_3}\right)^3 P_4
\ee
where subindex $3$ indicates the recircularized values before the accretion starts and subindex $4$ indicates the present values.
Assuming that the black hole accreted $5\msun$ from its companion after its formation, one can obtain
\be
P_3
   &=&\left(\frac{15.65\msun\times 70\msun}{10.65\msun\times 75\msun}\right)^3 3.45\ {\rm days} = 8.9\ {\rm days}
\ee
or a spin parameter of $a_\star\sim0.05$.  This, of course, obligates us to reconstruct our calculation in eq.~(\ref{P1-P2}), nevertheless, the preexplosion period is no longer restricted by the present-day spin parameter and the mass loss can be much smaller as we discuss below.

The black-hole progenitor star had to be more massive than the secondary star in order for it to evolve into a black hole at least a few million years before the black hole formation and accrete hypercritically $\sim5\msun$ after the black hole formation so we could observe the present-day configuration of the system.  Given such massive stars, we know that the mass loss through winds has to be considerable. So we know that the ZAMS mass of the black-hole progenitor had to be larger than that of the secondary star, and that the ZAMS mass of the secondary star had to be larger than its mass anytime afterwards, i.e.
\be
& M_{\rm ZAMS}>m_{\rm ZAMS}>m_{\rm pre\ explosion} \gtrsim m_{\rm after\ explosion} & \nonumber\\
& \gtrsim m_{\rm before\ accretion} > m_{\rm after\ accretion} + 5\msun &
\ee
where $M$ denotes the mass of black-hole progenitor and $m$ denotes the mass of the secondary star.
We have assumed that the mass of the secondary only changes drastically when it fills its Roche lobe for the first time (the second time will occur when it becomes a red giant, but the amount of mass transfer is much smaller) after the explosion of the primary star as explained by \citet{Pod03}: {\it ``After the brief turn-on phase, mass transfer occurs initially on the thermal timescale of the envelope reaching a peak mass-transfer rate of $\sim4\times10^{-3}\msun$yr$^{-1}$."}, at which point it transfers hypercritically and in a conservative way, i.e., with little or no mass loss, $5\msun$ to the black hole.  This means the ZAMS mass of the black-hole progenitor should be around $90\msun$.  And probably between 10 and $35\msun$ right before the explosion, depending on the intensity of the winds (see \citet{Bro01}).  This means that $\Delta M$ in equation~\ref{P1-P2} must be between 0 and $25\msun$. So,
\be
P_1=\left(1+\frac{\Delta M}{10.65+75}\right)^{-2}8.9\ {\rm days}
\ee
implies $5.3$ days $\le P_1\le 8.9$ days, or a natal spin parameter in the $0.05-0.1$ range for the black hole.

This result shows that the energy available for the Blandford-Znajeck mechanism to produce an explosion is very small. Following \citet{Lee00}, the black hole spin energy which can be extracted is given as,
\be
E_{BZ}=1.8\times10^{54}\epsilon_\Omega f(a_\star)\left(\frac{M}{\msun}\right) {\rm ergs}
\ee
where
\be
f(a_\star)=1-\sqrt{\frac{1}{2}\left[1+\sqrt{1- a_\star^2}\right]}.
\ee
Here
$
\epsilon_\Omega=\Omega_F/\Omega_H
$
is the efficiency of extracting rotational energy which, for an optimal process is $\sim0.5$, where $\Omega_F$ being the angular velocity of the magnetic field, and $\Omega_H$ the angular velocity of the black hole.
We obtain ``only" (as compared with the hundreds of bethes available in the Galactic transient sources \citep{BLMM07}) between 3 to 11 bethes of available energy.

This means that most likely the explosion was a dark one and the amount of expelled material was small if not zero,
analogous to that proposed by \citet{Mir03} for Cyg X$-$1.
\section{Conclusions}\label{Conc}

In this letter, we discussed a few problems were the currently observed spin of M33 X-7 all natal. Firstly the black hole progenitor overlap with the companion star, and secondly the binary will break up during the explosion in which the black hole was born.

We suggest that the hypercritical accretion has happened in M33 X-7 after the black hole formation which spins up the black hole after its formation.
M33 X$-$7 is the ideal system to test hypercritical accretion on.  Since $\dot M_{Edd}\equiv L_{Edd}/c^2=4\times10^{-8}\msun$yr$^{-1}$ the necessary accretion to have the black hole torqued up by its companion requires $\sim 120$ million years to acheive the $5\msun$ necessary to spin the black hole up to a spin of $a_\star=0.77$ \citep{KK99} as we suggested.  The age of the system is however only 2 to 3 million years \citep{Oro07}.

We think that hypercritical accretion was already established in \citet{Hou91}, \citet{Bro94} and \citet{Che95} by the disappearance of SN 1987A. However, it is good to have another proof as given by M33 X-7.

\acknowledgments
G.E.B. was supported by the US Department of Energy under Grant No. DE-FG02-88ER40388.  C.H.L. and I.H.P. were supported by Creative Research Initiatives (MEMS Space Telescope) of MEST/KOSEF.



\end{document}